\documentclass[aps,prb,twocolumn,superscriptaddress]{revtex4-2}
\usepackage{graphicx}
\usepackage{picture}
\usepackage{xcolor}
\usepackage{color}
\usepackage{tikz}
\usepackage{ulem}
\usepackage{amsmath}
\usepackage{rotating}
\usepackage{comment}

\begin{document}
\title{Tunable absorption spectrum splitting in a pulse-driven three-level system}

\author{Jiawei Wang}
\affiliation{Department of Physics, University at Buffalo SUNY, Buffalo, New York 14260, USA}
\author{Anthony Gullo}
\affiliation{Department of Physics, University at Buffalo SUNY, Buffalo, New York 14260, USA}
\author{Kavya Velmurugan} 
\affiliation{College of Computing, Georgia Institute of Technology, Atlanta, Georgia 30332, USA}
\author{Herbert F Fotso}
 \email{hffotso@buffalo.edu}
\affiliation{Department of Physics, University at Buffalo SUNY, Buffalo, New York 14260, USA}

\begin{abstract}

When a two-level system is driven on resonance by a strong incident field, its emission spectrum is characterized by the well-known Mollow triplet. 
If the absorption from the excited state, in this continuously driven two-level system, to a third, higher energy level, is probed by a weak field, the resulting absorption spectrum features the Autler-Townes doublet with two peaks separated by the Rabi frequency of the strong driving field. 
%
It has been shown that when the two-level system is instead driven by a periodic pulse sequence, the emission spectrum obtained has similarities with the Mollow triplet even though the system is only driven during the short application time of the pulses and is allowed to evolve freely between pulses. Here, we evaluate the absorption spectrum  of the three-level system in the ladder/cascade configuration when the bottom two levels are driven by a periodic pulse sequence while the transition between the middle and the highest level is probed by a weak field. The absorption spectrum displays similarities with the Autler-Townes doublet with frequency separation between the main peaks defined by the inter-pulse delay. In addition, this spectrum shows little dependence on the pulse carrier frequency. These results demonstrate the capacity to modulate the absorption spectrum of a three-level system with experimentally achievable pulse protocols. 





\end{abstract}

\maketitle

\section{Introduction}
\label{sec:introduction}

\noindent Driving quantum emitters with external control fields can drastically modify their spectral properties and, importantly, enable novel applications. For a two-level system, a strong resonant field produces the resonance fluorescence spectrum characterized by the Mollow triplet with a central peak at the emitter's frequency and two satellite peaks shifted by the Rabi frequency of the driving field on either side of the central peak~\cite{RF_Mollow_PhysRev1969, Mollow_PhysRevA_5_1972, Cohen_Tannoudji_Book1992}. Various studies have recently considered the possibility of replacing the continuous resonant driving field with a variety of pulse sequences and revealed a great deal of tunability in the resulting spectral properties~\cite{FotsoEtal_PRL2016, FotsoOtherPulsesJPhysB2018, vuckovic2020, IDS_2017}. For example, it was found that a periodic sequence of $\pi_x$ pulses applied on initially different or even noisy two-level systems can enhance the efficiency of Hong-Ou-Mandel two-photon interference experiments~\cite{Fotso_TPI_PRB_2019, Fotso_noisyTLS2022}. Similarly, it has been shown that the same sequence of $\pi_x$ pulses suppresses inhomogeneous broadening in the emission or absorption spectrum of a heterogeneous ensemble of two-level systems. The protocol refocuses the absorption spectrum of an ensemble of nitrogen vacancy centers and has thus been shown to restore the $1/T_2$ sensitivity when such an ensemble is used for magnetometry~\cite{FotsoDobrovitski_Absorption, JoasReinhard_BSensing}.

Beyond the two-level systems, when a third energy level is included, for a three-level system e.g. in the $\Lambda$ configuration, a driving field can give rise to electromagnetically induced transparency~\cite{Harris_EIT_1997, BollerImamogluHarrisPRL1991} with prominent applications in quantum memories~\cite{FleischhauerImamogluMarangosRevModPhys2005, Bajcsy_et_al_EIT_Nature2003, EIT_QtumMemory1, EIT_QtumMemory2, EIT_QtumMemory4}. The parent phenomenon, the Autler-Townes splitting, is observed for a strongly driven system and has also been implemented as a protocol for quantum memories~\cite{AutlerTownes1955, HeFiskMansonJApplPhys1992, PicquePinardJPhysB1976, SaglamyurekLeBlancNatPhot2018, RastogiLeBlancPRA2019}. Given the aforementioned effects of pulse sequences on two-level systems, it is then natural to investigate their effects on three-level systems and to explore potential advantages that they can provide in this context. In the present paper, we study the absorption spectrum  of the three-level system in the ladder or cascade configuration when the transition between the bottom two levels is driven by a periodic pulse sequence with inter-pulse delay $\tau$ while the transition between the middle and the highest energy level is probed with a weak field. We find that the resulting absorption spectrum displays similarities with the Autler-Townes doublet with separation between the main peaks defined by the inverse of the inter-pulse delay. In addition, this spectrum shows little dependence on the pulse carrier frequency. These results demonstrate a flexible capacity to modulate the absorption spectrum of a three-level system with achievable pulse protocols, highlighting alternative implementations of quantum memories in these systems. The pulse sequence produces an effect analogous to that of the continuously driven lower two levels with a Rabi frequency equal to $\pi/\tau$ while probing the transition between the first excited state and the highest excited state by a weak probing field. The findings in this paper are consistent with previous theoretical and experimental studies of pulse-induced tuning of spectral signatures of two and three-level systems\cite{YingyingEtAlPRA2022, YuxuanEtAlSciAdv2024, tanasittikosolPotvliegeArXiv2012}.

The rest of the paper is structured as follows. In section~\ref{sec:Model}, we introduce the model of the three-level system in the ladder configuration under the influence of the pulse sequence and the probing field. 
In this section, we also discuss the algorithm for our numerical solution of the relevant master equation governing the time evolution of the density matrix operator of the three-level system. In section~\ref{sec:Results}, we present the results of the spectrum and its dependence on the system and drive parameters before concluding the paper in section~\ref{sec:conclusion}.

\section{Model and Numerical Solution}
\label{sec:Model}
\noindent 

\begin{figure}[t] 
\includegraphics[width=8.40cm]{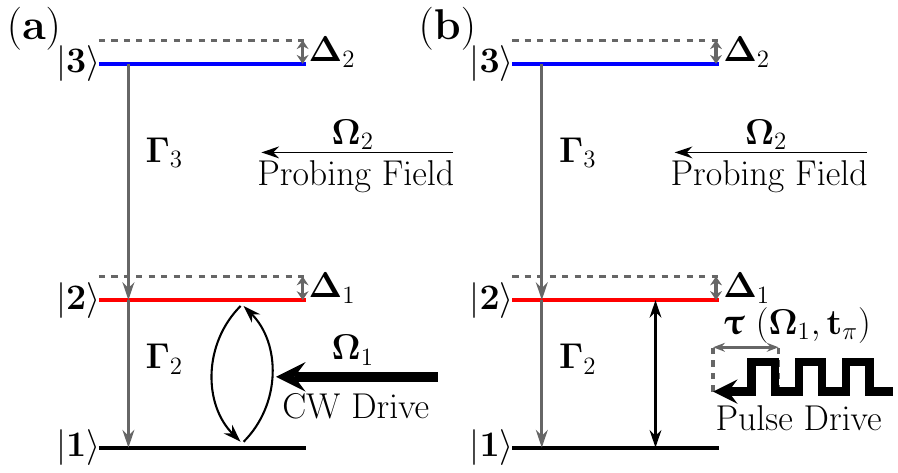}
\caption{Schematic representation of a three-level system in a ladder/cascade configuration. The Autler-Townes splitting is observed in the absorption from the first excited state $(|2\rangle)$ to the top excited state $(|3\rangle)$, probed by a weak field of Rabi frequency $\Omega_2$, when the bottom transition $(|1\rangle \leftrightarrow |2\rangle)$ is driven continuously on resonance by a strong continuous drive of Rabi frequency $\Omega_1$ (a). Here, we will evaluate the $(|2\rangle \leftrightarrow |3\rangle)$ absorption spectrum when the $(|1\rangle \leftrightarrow |2\rangle)$ is driven by a periodic sequence of $\pi$ pulses produced by periodically applying on the transition a field of Rabi frequency $\Omega_1$ for time $t_{\pi} = \pi/\Omega_1$ (b).}
\label{fig:Schematic}
\end{figure}

\noindent We consider a three-level system in the ladder configuration with energy levels $|1\rangle$, $|2\rangle$, and $|3\rangle$ with respective energies, $E_1$, $E_2$ and $E_3$ with $E_1 < E_2 < E_3$. This system is placed in a bosonic bath of photons to which it is coupled. In the usual formulation of the Autler-Townes effect, (Fig. \ref{fig:Schematic} (a)) the $|1\rangle \leftrightarrow |2\rangle$ transition is driven strongly  while the  $|2\rangle \leftrightarrow |3\rangle$ transition is gauged by a weak probing field. The measured absorption spectrum displays, when the driving field is resonant with the $|1\rangle \leftrightarrow |2\rangle$ transition, two symmetric peaks around the transition frequency that are separated by the Rabi frequency ($\Omega_1$) of the driving field.  This splitting is due to the dressed states, around the ground and the excited states, that are formed in the resonantly driven two-level system. A finite detuning ($\Delta_1$) between the driving field frequency $\nu_1$ and the transition frequency leads to an asymmetry in the position and the magnitude of the peaks.\\
Here, we consider the situation where the $|1\rangle \leftrightarrow |2\rangle$ transition is instead driven by a periodic sequence of finite width $\pi_x$ pulses (Fig. \ref{fig:Schematic} (b)). To this end, we apply a driving field with a time dependent Rabi frequency $\Omega_d(t)$ such that, $\Omega_d(t)= \Omega_1$ during the pulse and is zero otherwise. Similarly to the conventional Autler-Townes splitting setup, the probing field, at frequency $\hbar\omega = E_3 - E_2 + \Delta_2$ , with a Rabi frequency $\Omega_2 \ll \Omega_1$, is applied at all times. The driving field is non-zero for time $t_{\pi}=\pi/\Omega_1$. The relaxation rate from state $|2\rangle$ to state $|1\rangle$ is $\Gamma_2$ while the relaxation rate from state $|3\rangle$ to state $|2\rangle$ is $\Gamma_3$, and there is no direct transition allowed between states $|1\rangle$ and $|3\rangle$. We set $\Gamma_2 = 2$ and we measure our energy and time in units of $\Gamma_2/2$ and $2/\Gamma_2$ respectively.

The Hamiltonian describing the 3-level system in the radiation field, in the presence of the driving and probing field is made of five parts and can be written as:
\begin{equation}
    H = H_A + H_F + H_{AF} + H_d(t) + H_p(t).
    \label{eq:Hamiltonian}
\end{equation}
Where the individual terms of the Hamiltonian describe, respectively, the atomic energy levels ($H_A$), the radiation field modes ($H_{F}$), the coupling between the atom and the field mode ($H_{AF}$), the driving field on the $|1\rangle \leftrightarrow |2\rangle$ transition ($H_d(t)$), the probing field coupled to the $|2\rangle \leftrightarrow |3\rangle$ transition (($H_p(t)$)). These individual terms of the Hamiltonian in Eq.(\ref{eq:Hamiltonian}) are defined by:
\begin{eqnarray}
H_A &=& \hbar \omega_1 | 1 \rangle\langle 1 | + \hbar \omega_2 | 2 \rangle\langle 2 | + \hbar \omega_3 | 3 \rangle\langle 3 | ,\\
H_F &=& \sum_k \hbar \omega_k a_k^{\dagger} a_k ,\\
H_{AF} &=& i \sum_k g_{12}^k a^{\dagger}_k |1\rangle \langle 2 | + i \sum_k g_{13}^k a^{\dagger}_k |1\rangle \langle 3 |  \nonumber \\
 & & \;\;\;\;\;\;  \;\;\;\;\;\;  + i \sum_k g_{23}^k a^{\dagger}_k |2\rangle \langle 3 | + H.c. \\
H_{d}  &= & \frac{\Omega_d(t)}{2} \mathrm{e}^{-i \nu_1 t}| 1 \rangle\langle 2 | +  H.c. , \\
H_{p}  &=& \frac{\Omega_2}{2} \mathrm{e}^{-i \omega t} | 2 \rangle\langle 3 | +  H.c.,
\end{eqnarray}
$g^k_{12}, \; g^k_{13}, \; g^k_{23}$ are, respectively, the coupling strengths of the transition between state $|1\rangle$ and state $|2\rangle$, state $|1\rangle$ and state $|3\rangle$, and,  state $|2\rangle$ and state $|3\rangle$ with the radiation mode $k$. it is assumed that $g^k_{23}$ is negligible for all modes. The operators $a^{\dagger}_k$ and $a_k$ are, respectively, the creation and the annihilation operators for the $k$ radiation mode with frequency $\omega_k$. The relaxation rate $\Gamma_2$ of the $|2\rangle \to |1\rangle$ transition is given by $\Gamma_2 = 2\pi\int (g^k_{12})^2 \; \delta(\omega_k - (\omega_2 - \omega_1)) \; dk$ and similarly for $\Gamma_3$.
The driving field and the probing field are treated at a semi-classical level which is consistent with driving the emitter with a coherent field~\cite{Cohen_Tannoudji_Book1992}. 

To characterize the time evolution of the system, we will use the density matrix operator of the 3-level system when the bosonic degrees of freedom are integrated out of the problem. 
This density matrix operator of the 3-level system is given by: $\rho =  \rho_{11} | 1 \rangle\langle 1 |  +
  \rho_{22} | 2 \rangle\langle 2 |  +  
  \rho_{33} | 3 \rangle\langle 3 |   
   +  \rho_{21} | 2 \rangle\langle 1 |  + 
  \rho_{12} | 1 \rangle\langle 2 |  +
  \rho_{32} | 3 \rangle\langle 2 |   
   +  \rho_{23} | 2 \rangle\langle 3 |  +
  \rho_{31} | 3 \rangle\langle 1 |  +
  \rho_{13} | 1 \rangle\langle 3 |  $.

In general, $\rho_{ij}$'s for $i,j = 1,2,3$ are complex numbers. However, as a result of hermiticity, $\rho_{ii}$'s for $i=1,2,3$ are real and only 6 matrix elements of the density matrix operator are independent. Furthermore, we have the identity $\rho_{11} + \rho_{22} + \rho_{33} = 1$.\\
One can use the Lindblad equation or the independent-rate approach to obtain the equations of motion due to the different terms of the Hamiltonian. The resulting master equations, stated below term by term, characterize the time evolution of the density matrix operator~\cite{Cohen_Tannoudji_Book1992, MilonniKnight1980}.

\begin{eqnarray}
\dot{\rho}_{11} & = &  \Gamma_2 \rho_{22} -i \frac{\Omega_d(t)}{2} \rho_{21} + i \frac{\Omega_d(t)}{2}  \rho^*_{21} \label{eq:rho_1} \\
\dot{\rho}_{22} & = & -\Gamma_2 \rho_{22} + \Gamma_3 \rho_{33} + i \frac{\Omega_d(t)}{2} \rho_{21} \nonumber \\
& - & i \frac{\Omega_d(t)}{2} \rho^*_{21}  -i \frac{\Omega_2}{2} \rho_{32} + i \frac{\Omega_2}{2} \rho^*_{32}  \label{eq:rho_2} \\
\dot{\rho}_{33} & = &  -\Gamma_3 \rho_{33} +i \frac{\Omega_2}{2} \rho_{32} -i \frac{\Omega_2}{2} \rho^*_{32} \label{eq:rho_3} \\
\dot{\rho}_{21} & = & -i \frac{\Omega_d(t)}{2}\rho_{11} +  i \frac{\Omega_d(t)}{2} \rho_{22} \nonumber \\
& + & \left[ i \Delta_1 -\frac{1}{2}\Gamma_2 \right] \rho_{21} -i \frac{\Omega_2}{2}\rho_{31} \label{eq:rho_4} \\
\dot{\rho}_{32} & = & -i \frac{\Omega_2}{2} \rho_{22} +  i \frac{\Omega_2}{2} \rho_{33} \nonumber \\
& + & \left[ i \Delta_2 -\frac{1}{2}\Gamma_{32}\right] \rho_{32}  +i \frac{\Omega_d(t)}{2} \rho_{31} \label{eq:rho_5} \\
\dot{\rho}_{31} & = & -i \frac{\Omega_2}{2} \rho_{21}   +  i \frac{\Omega_d(t)}{2} \rho_{32} \nonumber \\
& + & \left[ i (\Delta_1 +\Delta_2) -\frac{1}{2}\Gamma_{3} \right]\rho_{31}
\label{eq:rho_6} 
\end{eqnarray}
Here factors of $\mathrm{e}^{i\nu_1 t}$, $\mathrm{e}^{i\omega t} $, and $\mathrm{e}^{i(\nu_1 + \omega)t} $ have been included in $\rho_{21}$, $\rho_{32}$ and $\rho_{31}$ respectively.
Note also that $\rho_{12}$ and $\rho_{23}$ have respectively been replaced in these equations by $\rho_{21}^*$ and $\rho_{32}^*$, using the above-mentioned properties of the density matrix operator. 
We have introduced the parameter $\Gamma_{32}= \Gamma_3 + \Gamma_2$.
For a continuously driven $|1\rangle \leftrightarrow |2\rangle$ transition, a steady expression for the occupation of the highest energy level has been derived~\cite{MilonniKnight1980}. For the pulse driven system, however, the solution becomes untractable and we resort to a numerical solution. 

At time $t = 0 $, the system is prepared in the state $| 2 \rangle$. All matrix elements of the density matrix operator are set to zero except for $\rho_{22} = 1$. The transition between the lower energy level $|1\rangle$ and the intermediate energy level $|2\rangle$ is then driven periodically with inter-pulse delay $\tau$, and pulses applied by a field of Rabi frequency $\Omega_1$, that is applied for a time $\pi/\Omega_1$ amounting to a finite width $\pi$-pulse. More explicitly, the evolution of the master equation (\ref{eq:rho_1}-\ref{eq:rho_6}) is carried out periodically with a $\tau$ time interval corresponding, first to an evolution with $\Omega_d(t) = 0$ for time $\tau - t_{\pi}$, followed, for time $t_{\pi}$ to an evolution with $\Omega_d(t) = \Omega_1$.  
The weak probing field, with Rabi frequency $\Omega_2 (\ll \Omega_1)$ is applied at all times on the transition between the intermediate energy level $|2\rangle$ and the upper energy level $|3\rangle$. 

\noindent We integrate these differential equations iteratively starting from the initial conditions for the density matrix operator $\rho(t_0 = 0)$ and calculate $\rho(t_0 + \delta t) $, $\rho(t_0 + 2 \delta t)$, ... up to $\rho(t_0 + N_t \delta t) = \rho(t_{max})$. Where $t_{max} = N_t \delta t$ and $N_t$ is the number of time steps of width $\delta t$. The time step $\delta t$ is chosen such that the discretization error is negligible.
The time evolution of the density matrix characterized by equations (\ref{eq:rho_1} - \ref{eq:rho_6}), is carried out for frequencies $\omega$ of the weak probing field, covering a range of values of $\Delta_2$ from $\omega_{min} = -40$ to $\omega_{max} = 40$ outside of which there is negligible spectral weight.
For each probing field frequency, we evolve the system up to a maximum time when the system has reached the steady state. The steady state here is identified by tracking the time evolution of the density matrix operator, particularly the occupation number of the highest excited state, $\rho_{33}$, and confirming that it is no longer changing appreciably.

\begin{figure}[t] 
\includegraphics[width=8.40cm]{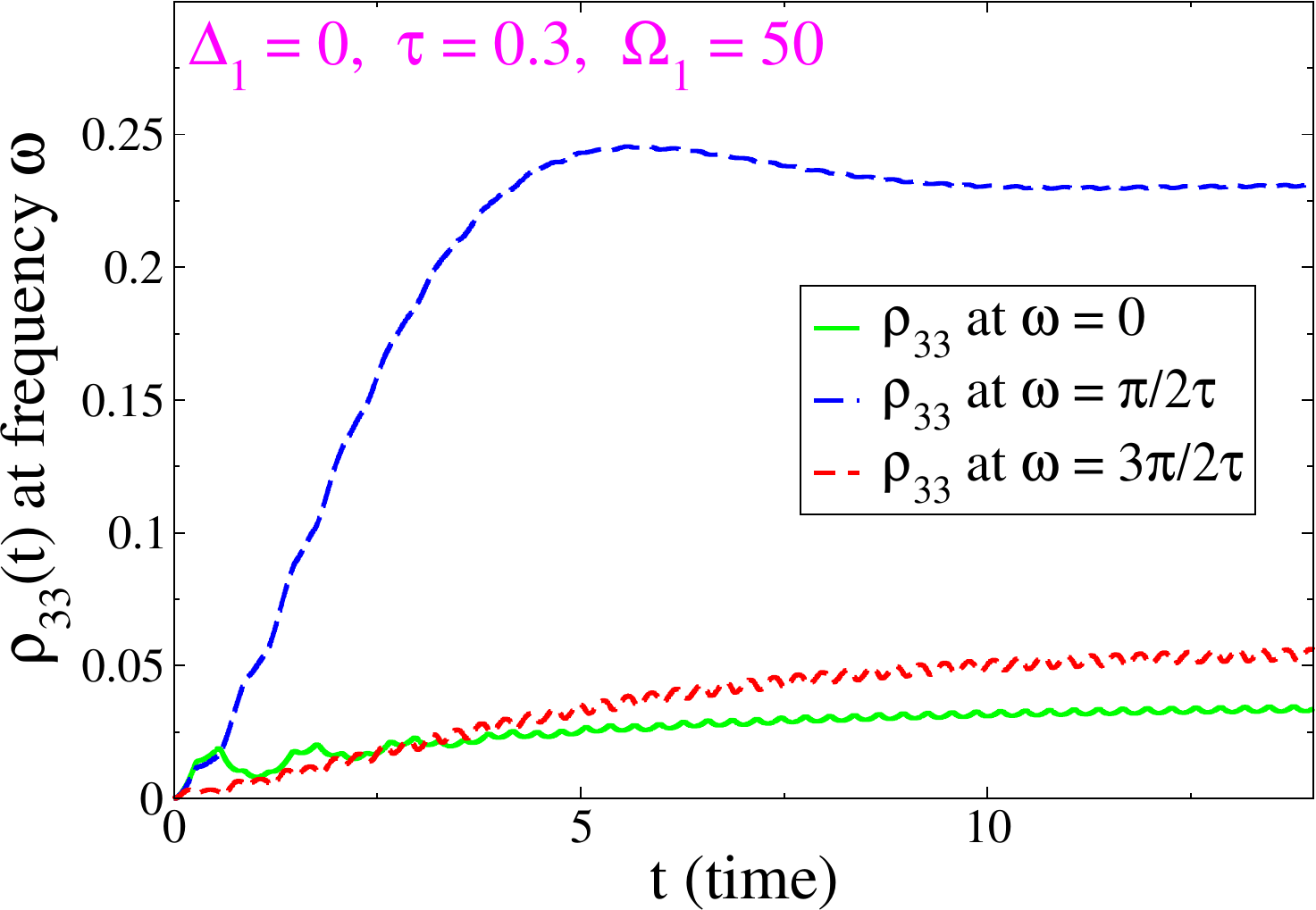}
\caption{Population of the second excited state ($|3\rangle)$ as a function of time, $\rho_{33}(t)$, when the $(|1\rangle \leftrightarrow |2\rangle)$ transition is driven by a periodic sequence of $\pi$ pulses with period $\tau = 0.3$. The pulse carrier frequency is resonant with the transition, $\Delta_1 = 0$, and the pulses are due to a driving field of Rabi frequency $\Omega_1 = 50$ that is periodically applied for time $t_{\pi}=\pi/\Omega$. $\rho_{33}(t)$ is shown for $\omega = 0$ (solid green line), $\pi/2\tau$ (dashed blue line) and $3\pi/2\tau$ (dashed red line). Time is measured in units of $2/\Gamma_2$}
\label{fig:rho33_vs_time}
\end{figure}

The absorption of the probing field, of interest in the present study, is proportional to the population of the energy level $|3\rangle$. Thus, it is studied through the occupation of the highest excited state, $\rho_{33}$, in this steady state, for varying values $\omega$ of the probing field frequency. 
The population of the highest energy level is probed within our weak probe treatment and in the steady state. The lineshape in this long-time regime is found, within the weak probe regime, to not depend on the actual value of the probing field Rabi frequency as long as it does not significantly modify the dynamics. However, the observed population of the highest energy level, in its magnitude, may have some dependence on the probing field Rabi frequency.

\begin{figure} [t] 
\includegraphics[width=8.40cm, height=6.0cm]{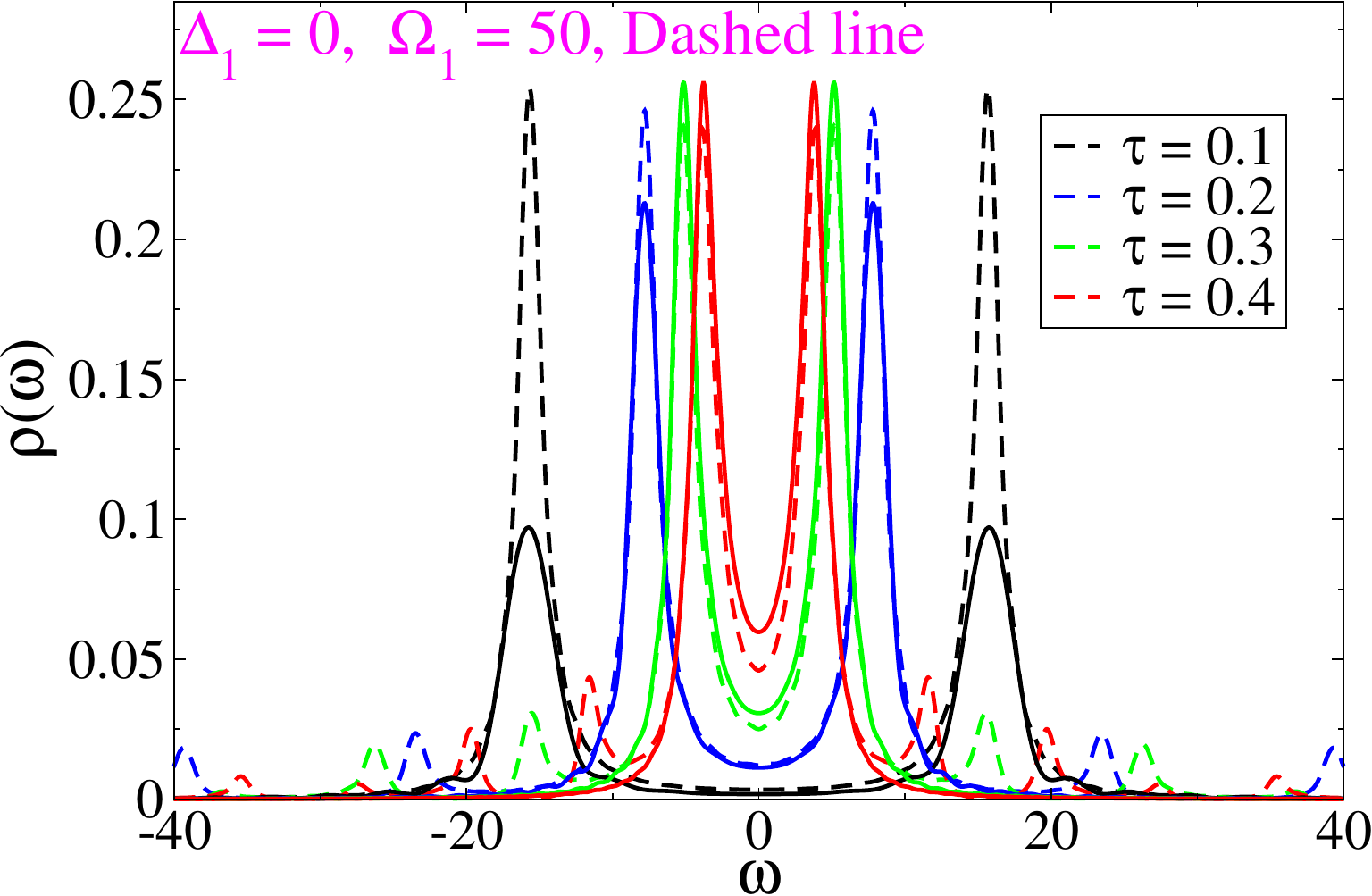}
\caption{Population of the second excited state or absorption spectrum of the $(|2\rangle \leftrightarrow |3\rangle)$ transition when the $(|1\rangle \leftrightarrow |2\rangle)$ is driven by a periodic sequence of $\pi$ pulses with period $\tau$ the pulse is due to a driving field of Rabi frequency $\Omega_1 = 50$ that is applied for time $t_{\pi}=\pi/\Omega_1$. Absorption spectrum shown for detuning $\Delta_1 = 0$ under pulse sequences of different periods: $\tau = 0.1$ (black dashed line), $\tau = 0.2$ (dashed blue line), $\tau = 0.3$ (dashed green line), $\tau = 0.4$ (dotted red line). The solid lines show the continuously driven system with frequency $\pi/\tau$ and all other parameters identical to the pulse-driven case. All energies are measured in units of $\Gamma_2/2$.}
\label{fig:Delta0p0_4Taus}
\end{figure}

\begin{figure}[h]
    \centering
    \includegraphics[width=8.20cm, height=5.80cm]{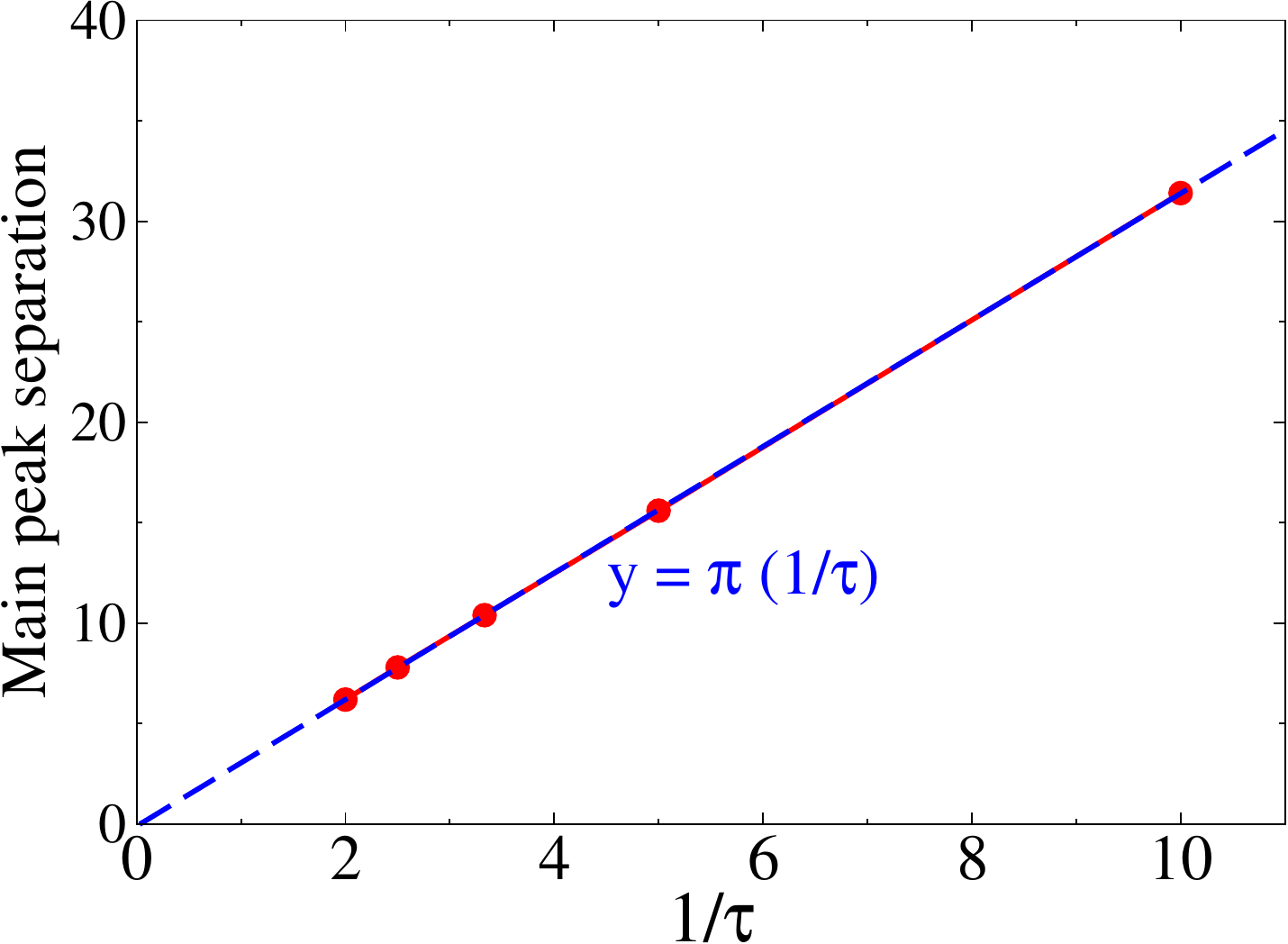}
    \caption{Separation between the main peaks as a function of $1/\tau$ for $\Delta_1 = 0$. The red line and symbols represent the data while the dashed blue line represents the linear fit with slope $\pi$ confirming the dependence on $\tau$. All energies are measured in units of $\Gamma_2/2$.}
    \label{fig:peakSeparation}
\end{figure}

\begin{figure}[t] 
\includegraphics[width=8.20cm]{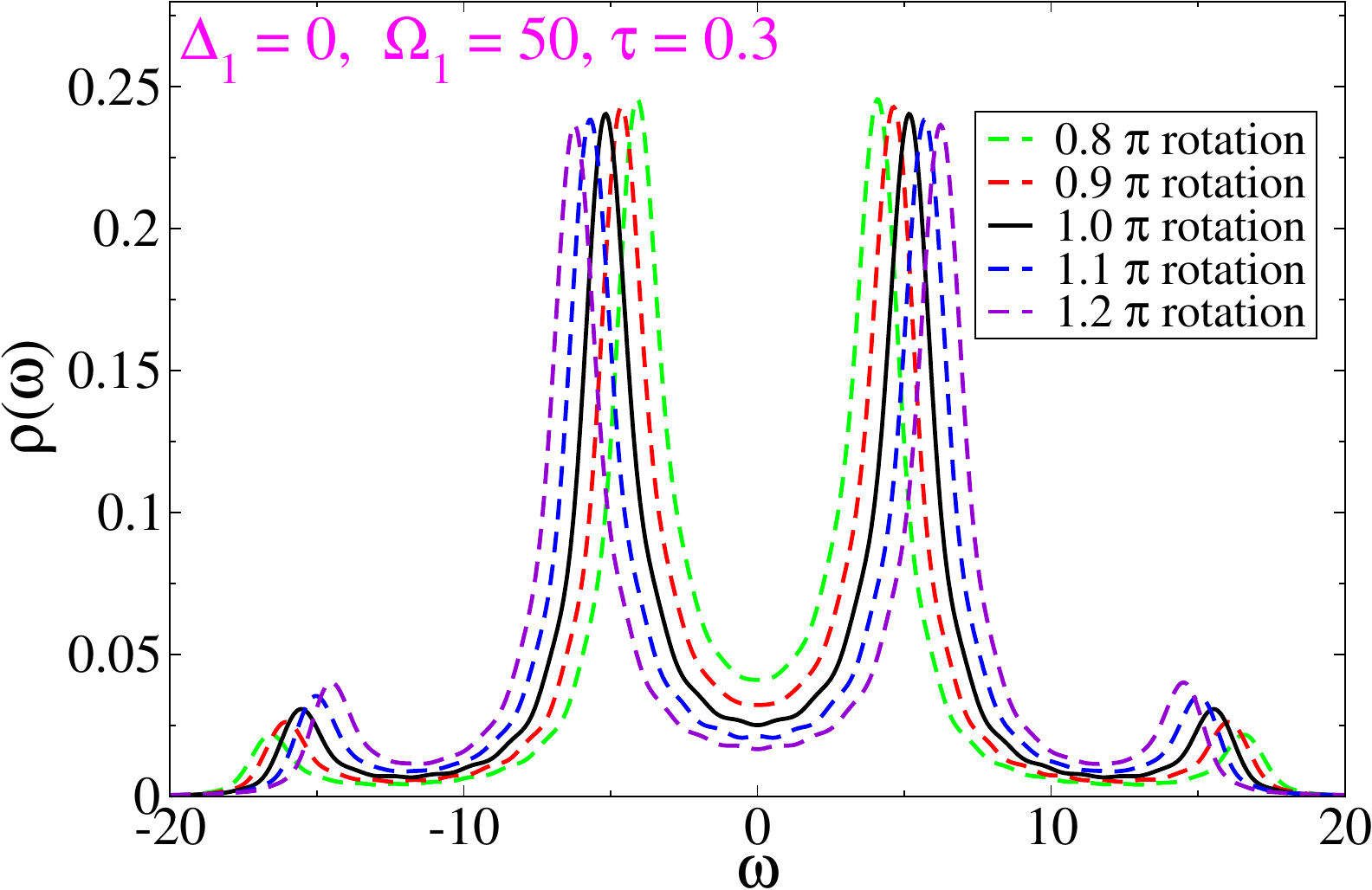}
\caption{Population of the second excited state or absorption spectrum of the $(|2\rangle \leftrightarrow |3\rangle)$ transition when the $(|1\rangle \leftrightarrow |2\rangle)$ is driven by a periodic sequence of imperfect pulses accomplishing a $0.8 \pi $ rotation (dashed green line), $0.9 \pi $ rotation (dashed red line), $1.0 \pi $ rotation (solid black line), $1.1 \pi $ rotation (dashed blue line), $1.2 \pi $ rotation (dashed purple line),  with period $\tau = 0.3$ and $\Omega_1 = 50$. The detuning with the pulse carrier frequency is $\Delta_1 = 0$.  All energies are measured in units of $\Gamma_2/2$.}
\label{fig:Tau0p3_ApproxPi_X5}
\end{figure}

\begin{figure}[t] 
\includegraphics[width=8.20cm]{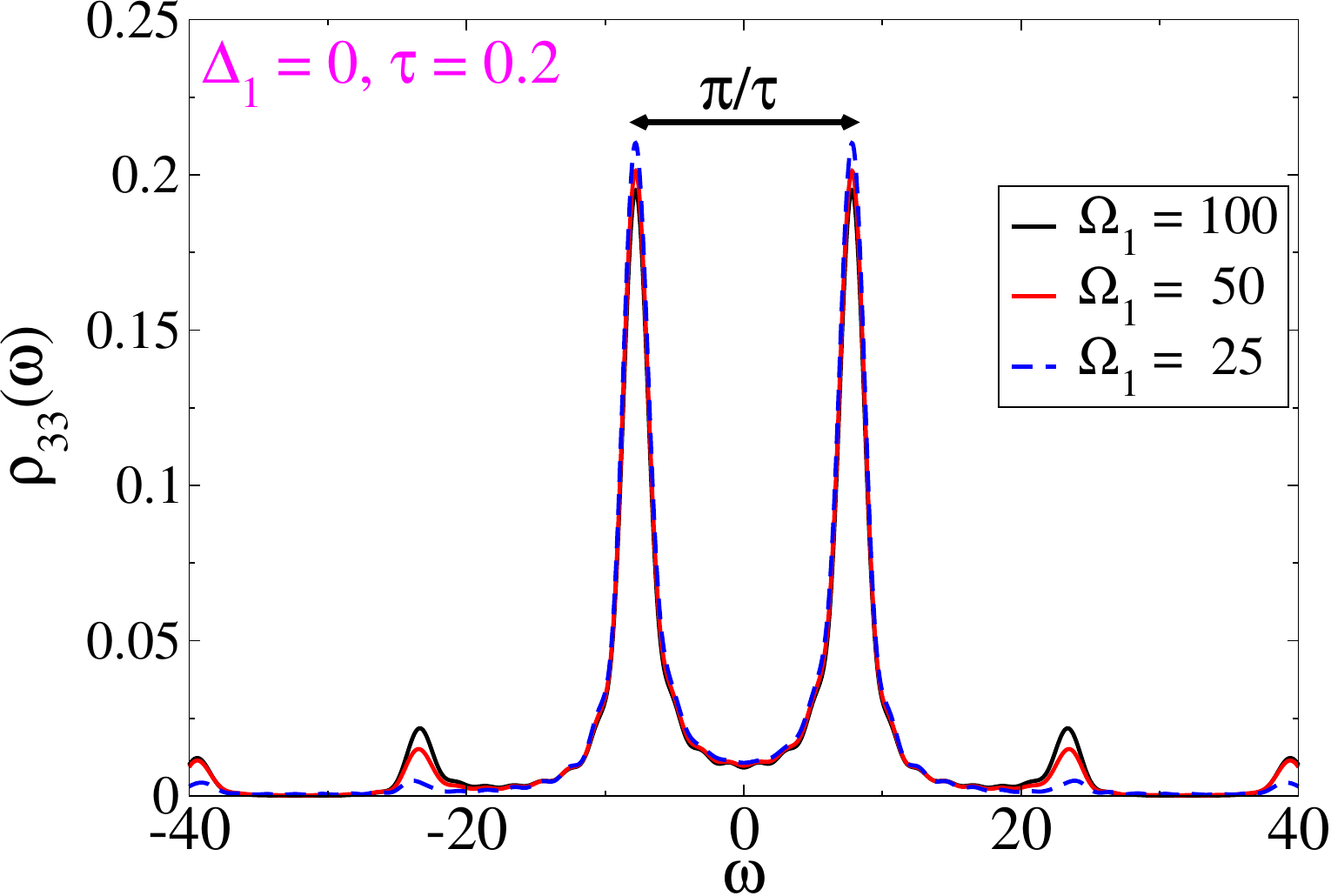}\\
\vspace{3mm}
\includegraphics[width=8.20cm]{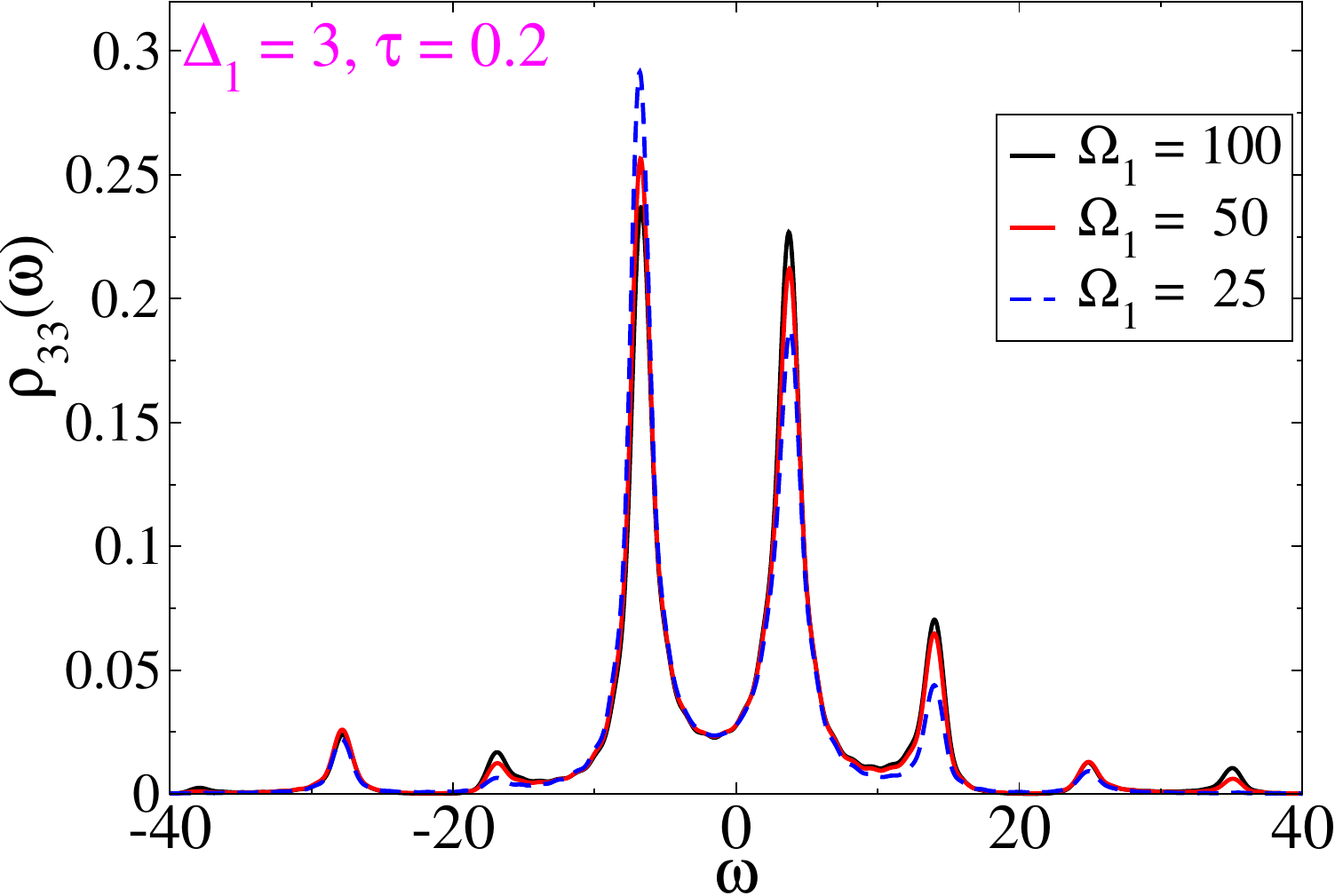}
\caption{Population of the second excited state or absorption spectrum of the $(|2\rangle \leftrightarrow |3\rangle)$ transition when the $(|1\rangle \leftrightarrow |2\rangle)$ is driven by a periodic sequence of $\pi$ pulses with period $\tau$ the pulse is due to a driving field of Rabi frequency $\Omega_1 = 100$ (solid back line), $\Omega_1 = 50$ (solid red line), $\Omega_1 = 25$ (dashed blue line), that is applied for time $t_{\pi}=\pi/\Omega_1$. The pulse period is $\tau = 0.2 $ and the detuning is $\Delta_1 = 0$ (top) and $\Delta_1 = 3$ (bottom). All energies are measured in units of $\Gamma_2/2$.}
\label{fig:Delta0_Delta3p0_3Omegas}
\end{figure}

\begin{figure}[t] 
\includegraphics[width=8.40cm]{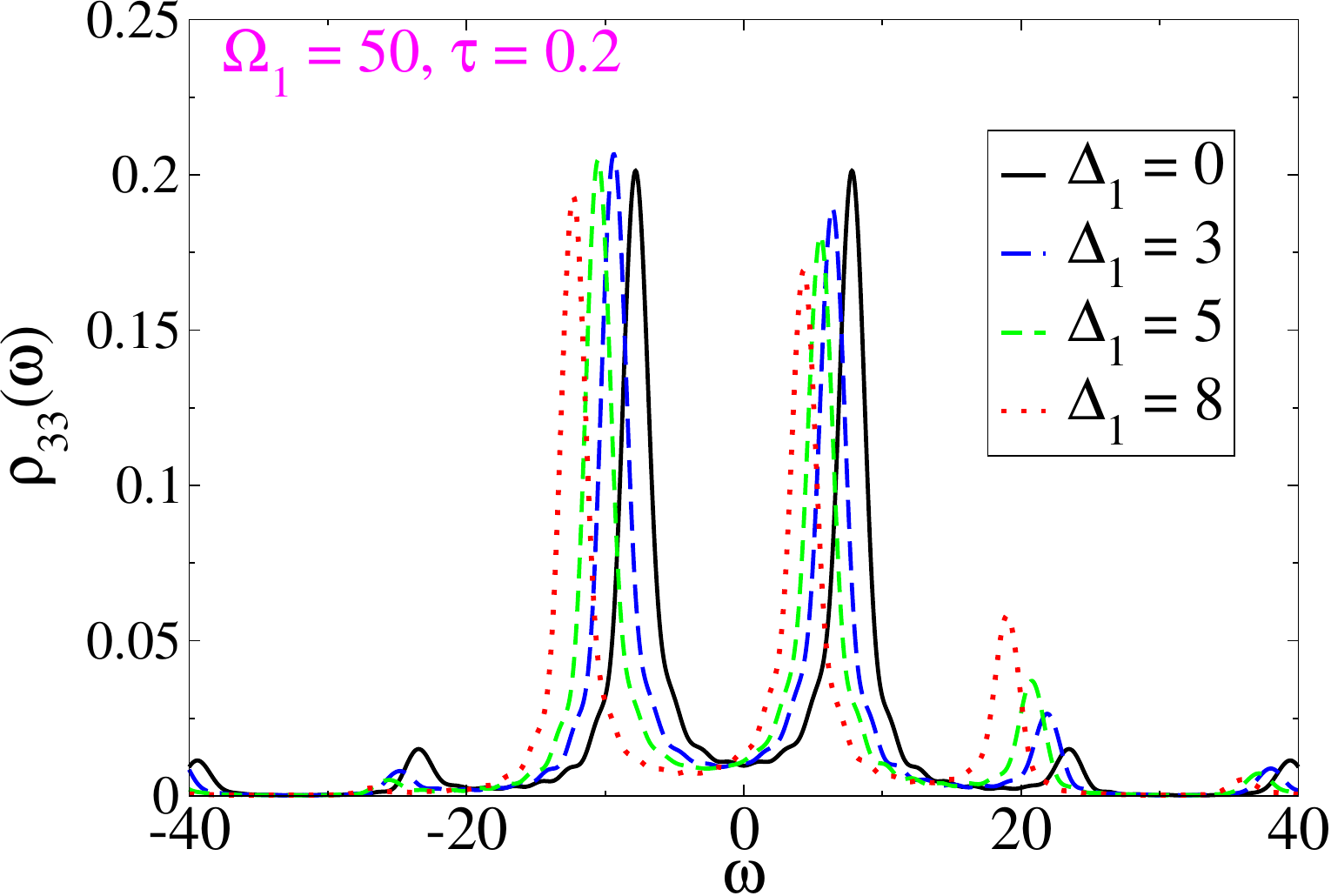}
\vspace{3mm}
\includegraphics[width=8.20cm]{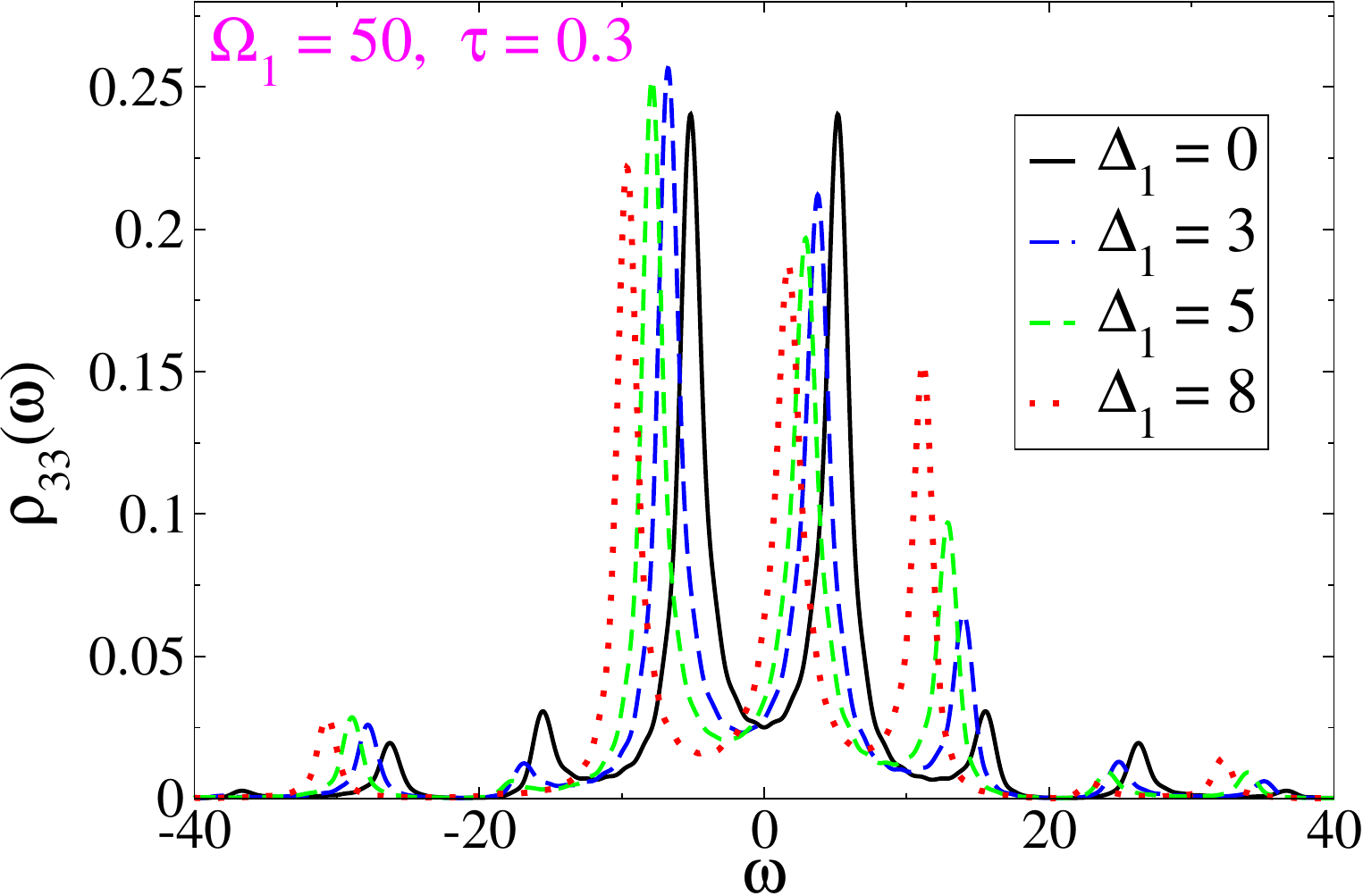}
\caption{Population of the second excited state or absorption spectrum of the $(|2\rangle \leftrightarrow |3\rangle)$ transition when the $(|1\rangle \leftrightarrow |2\rangle)$ is driven by a periodic sequence of $\pi$ pulses with period $\tau = 0.2$ (top) and $\tau = 0.3$ (bottom). The $\pi$ pulses are achieved with Rabi frequency $\Omega_1 = 50$. The detuning with the pulse carrier frequency is $\Delta_1 = 0$ (solid black line), $\Delta_1 = 3$ (dashed blue line), $\Delta_1 = 5$ (dashed green line), $\Delta_1 = 8$ (dotted red line). All energies are measured in units of $\Gamma_2/2$.}
\label{fig:Tau0p2_4Deltas}
\end{figure}

\section{Results}
\label{sec:Results}
\noindent We start by assessing the realization of the steady state in our pulse-driven system. Fig.~\ref{fig:rho33_vs_time} shows $\rho_{33}(t)$, the population of the highest excited state ($|3\rangle)$ as a function of time,  when the $(|1\rangle \leftrightarrow |2\rangle)$ transition is driven by a periodic sequence of $\pi$ pulses with period $\tau = 0.3$. Here, the pulses are due to a driving field of Rabi frequency $\Omega_1 = 50$ that is applied for time $t_{\pi}=\pi/\Omega_1$. $\rho_{33}(t)$ is shown for $\omega = 0$ (solid green line), $\pi/2\tau$ (dashed blue line), $3\pi/2\tau$ (dashed red line). The figure clearly shows that after an initial non-trivial transient, the system settles into a steady state where the population for any frequency does not show any non-trivial change as a function of time. The rest of the data that we present throughout this paper corresponds to this steady state. Here and in what follows, we choose $\Gamma_3 = 0.2$ and $\Omega_2 = 1.0$ but we note that the same lineshape is obtained for $\Omega_2 = 0.1$ with the only difference that the peaks in the absorption spectrum have higher magnitudes for larger $\Omega_2$. 

Fig.~\ref{fig:Delta0p0_4Taus} shows, for detuning $\Delta_1 = 0$, the absorption spectrum of the $(|2\rangle \leftrightarrow |3\rangle)$ transition, or the population of the highest excited state as a function of frequency, when the $(|1\rangle \leftrightarrow |2\rangle)$ is driven by a periodic sequence of $\pi$ pulses with different values of the period $\tau$. The pulse is due to a driving field with Rabi frequency $\Omega_1 = 50$ that is applied for time $t_{\pi}=\pi/\Omega_1$. The figure shows results for inter-pulse delays, $\tau = 0.1$ (black dashed line), $\tau = 0.2$ (dashed blue line), $\tau = 0.3$ (dashed green line), $\tau = 0.4$ (dotted red line).
The spectrum features two symmetric main peaks separated by a frequency $\pi/\tau$ with a clear similarity with the Autler-Townes splitting obtained for a continuous resonant drive~\cite{AutlerTownes1955, AutlerTownes1, MilonniKnight1980}. The separation between the main peaks is shown in Fig.~\ref{fig:peakSeparation} that presents this separation between the main peaks as a function of $1/\tau$. The figure clearly shows that this separation in the pulse-driven system is equal to $\pi/\tau$.  Note that there are additional peaks centered around frequencies that are integer multiples of $\pm \pi/2\tau$ but with exponentially suppressed spectral weights. These additional peaks with smaller spectral weight are visible in the figure for large $\tau$ values. The solid lines in Fig.~\ref{fig:Delta0p0_4Taus} show the absorption of the continuously driven system with frequency $\pi/\tau$ and all other parameters identical to the pulse-driven case for $\tau = 0.1$ (black line), $\tau = 0.2$ (blue line), $\tau = 0.3$ (red line), $\tau = 0.4$ (green line). Note the identical peak locations. The peak heights are different because of the relative values of the probing field Rabi frequency with respect to the driving field.

We show the requirement of the $\pi$ pulses in Fig.~\ref{fig:Tau0p3_ApproxPi_X5} where the absorption is shown for a periodic sequence of pulses with each pulses accomplishing a $0.8 \pi $ rotation (dashed green line), $0.9 \pi $ rotation (dashed red line), $1.0 \pi $ rotation (solid black line), $1.1 \pi $ rotation (dashed blue line), $1.2 \pi $ rotation (dashed purple line),  with period $\tau = 0.3$ and $\Omega_1 = 50$. The detuning with the pulse carrier frequency is $\Delta_1 = 0$. We see that as the pulses are tuned away from a $\pi$ rotation, the peak separations are adjusted away from the $\pi/\tau$ value.

Next, we examine the dependence of this $|2\rangle \leftrightarrow |3\rangle$ absorption spectrum on the Rabi frequency of the pulses. This is equivalent to examining how this spectrum varies with the duration of the $\pi$ pulses. Fig.~\ref{fig:Delta0_Delta3p0_3Omegas} shows this absorption spectrum for a pulse period
$\tau = 0.2$ for $\Delta_1 = 0$ (top) and $\Delta_1 = 3$ (bottom) for pulse Rabi frequency $\Omega_1 = 100$ (solid black line), $\Omega_1 = 50$ (solid red line), $\Omega_1 = 25$ (dashed blue line). One readily observes a weak dependence on the Rabi frequency. The lineshape is preserved even for relatively broad pulses ($t_{\pi} \sim \tau/2$). For finite detuning, we observe an asymmetry in the absorption spectrum and with more spectral weight shifted to the satellite peaks. This feature is emphasized in Fig.~\ref{fig:Tau0p2_4Deltas} that shows the absorption spectrum for different detunings. Here, $\tau = 0.2$ (top panel) and $\tau = 0.3$ (bottom panel). The $\pi$ pulses are achieved by a driving field with Rabi frequency $\Omega_1 = 50$. The detuning with the pulse carrier frequency is $\Delta_1 = 0$ (solid black line), $\Delta_1 = 3$ (dashed blue line), $\Delta_1 = 5$ (dashed green line), $\Delta_1 = 8$ (dotted red line). Here, we observe the shift in the location of the peaks and their relative amplitudes as a function of the detuning with more pronounced asymmetry found in the results for larger inter-pulse delay.

\section{Conclusion}
\label{sec:conclusion}
The Autler-Townes effect and its parent phenomenon, electromagnetically induced transparency, have been actively investigated since their discoveries. Their applications have been demonstrated in recent advances as quantum memories in quantum information processing. The Autler-Townes effect is readily explained by the dressed state picture that is also responsible for the Mollow triplet of resonance fluorescence occuring when a two-level system is driven on resonance by a strong incident field.
Previous work has shown that when a two-level system is instead driven by a periodic sequence of $\pi$ pulses, the emission spectrum obtained has similarities with the Mollow triplet. This spectrum features a central peak at the pulse carrier frequency and two satellite peaks at $\pm \pi/\tau$ where $\tau$ is the inter-pulse delay of the sequence. In the present paper, we have evaluated, the absorption spectrum of the three-level system in the ladder configuration when the bottom two levels are driven by a periodic sequence of $\pi$ pulses, while the transition between the middle and the highest level is probed by a weak field.
For a pulse carrier frequency that is resonant with the transition frequency of the bottom two levels, this absorption spectrum displays similarities with the Autler-Townes doublet. It features two main peaks at frequencies separated by $\pi/\tau$. This spectrum is found to have little dependence on the pulse carrier frequency and shows little change even for rather broad pulses (pulse duration $\sim \tau/2$). For finite detuning, the spectrum has an asymmetry that increases with the detuning. These results demonstrate the capacity to modulate the absorption spectrum of a three-level system with experimentally achievable pulse protocols and provide an alternative pathway for realizing quantum memories in these systems.

\section*{Acknowledgment} 
\noindent We acknowledge support from the National Science Foundation under Grants No. PHY-2014023 and No. QIS-2328752.



\bibliography{mainBib}





\end{document}